# Optimal confidence intervals for bounded parameters

## A correct alternative to the recipe of Feldman and Cousins

Fyodor V. Tkachov
Institute for Nuclear Research of the Russian Academy of Sciences
Moscow 117312, Russian Federation
ftkachov@ms2.inr.ac.ru

**Abstract.** A priori bound for the parameter to be estimated is incorporated into confidence intervals within frequentistic approach in a straightforward and optimal fashion, ensuring the best resolution of non-boundary values as well as robustness for non-physical values of the estimator.

Consider the problem of how the conventional confidence intervals are to be modified in order to incorporate the a priori information that the estimated parameter is bounded, e.g. $\theta \geq 0$. An immediate motivation is provided by the neutrino mass measurement experiments [1].

The problem was addressed in ref. [2] that has accumulated a thousand citations, which fact emphasizes the importance of the problem. However, instead of properly solving the problem, ref. [2] applied a magic patch in the form of a ratio of likelihoods — even if it was done within the freedom allowed by Neyman's construction of confidence intervals. As a result, the recipe of ref. [2] is unsatisfactory in that it may yield vanishing confidence intervals for large negative measured values of the parameter $\theta$ where one ought to worry about one's theoretical model of the experiment rather than celebrate tight bounds on $\theta$.

The purpose of this letter is to present a straightforward, purely frequentistic solution without invoking any magic. The solution is optimal in two respects:

1) it resolves non-boundary values from the boundary in the best possible way;

2) it avoids the artifact of vanishing confidence intervals for deeply non-physical values.

The solution is rather general: it is valid in any situation where the conventional confidence intervals can be constructed. The final recipe consists in a simple modification of the conventional confidence belt (i.e. the one constructed without regard for the inequality) for the confidence level $\beta$. The additional pieces come from another such confidence belt, the one for the confidence level $\tilde{\beta} = 1 - 2(1-\beta) < \beta$ (see Fig. 6).

## 1. Setup

Let $\hat{\theta}$ be a conventional estimator for the unknown parameter $\theta$, i.e. an esimator constructed without regard for the a priori bound (e.g. obtained via the paradigmatic method of moments [3], [4]).

The random variable $\hat{\theta}$ is a function of a set of experimental data $X$: $\hat{\theta} = \hat{\theta}(X)$. Its probability density $d_\theta(\hat{\theta})$ is parameterized by $\theta$ and is assumed to be known and non-singular as required in the standard Neyman's construction of confidence intervals (see below). The density $d_\theta(\hat{\theta})$ incorporates all the information about the experiment (including the estimation method) in regard of the measurement of $\theta$.

Let $\alpha$, $\alpha'$ be small and non-negative. Define $L_\alpha(\theta)$ and $U_{\alpha'}(\theta)$ according to

$$P\left(-\infty < \hat{\theta} < L_\alpha(\theta)\right) = \alpha, \quad P\left(U_{\alpha'}(\theta) < \hat{\theta} < +\infty\right) = \alpha' \tag{1}$$

Verbally: the probability for the estimator to fall below $L_\alpha(\theta)$ is $\alpha$; above $U_{\alpha'}(\theta)$, $\alpha'$.
The $L_\alpha(\theta)$ thus defined corresponds to the $Z_\alpha$ defined in sec. 9.1.1 of [3].

Assuming $L_\alpha(\theta)$ and $U_{\alpha'}(\theta)$ to be invertible functions of $\theta$, eqs. (1) can be rewritten as follows:

$$P\left(l_\alpha(\hat{\theta}) < \theta\right) = \alpha, \quad P\left(\theta < u_{\alpha'}(\hat{\theta})\right) = \alpha' \tag{2}$$

where $u_\alpha = U_\alpha^{-1}$, $l_\alpha = L_\alpha^{-1}$. Eqs. (2) say that the probabilities for the random variables $l_\alpha(\hat{\theta})$, resp. $u_{\alpha'}(\hat{\theta})$ to fall below, resp. above the unknown true value $\theta$, are $\alpha$, resp. $\alpha'$.

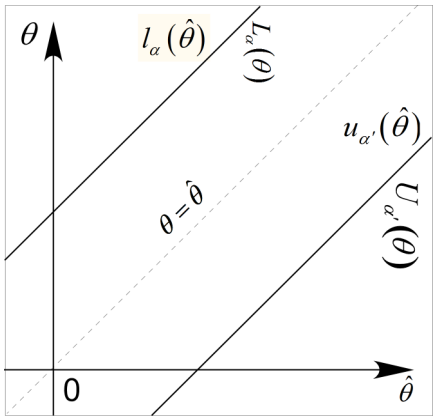

**Fig. 1.** Shown are the functions $\theta = l_\alpha(\hat{\theta})$ and $\theta = u_{\alpha'}(\hat{\theta})$ (or $\hat{\theta} = L_\alpha(\theta)$ and $\hat{\theta} = U_{\alpha'}(\theta)$, depending on the viewpoint).
In general, the diagonal $\theta = \hat{\theta}$ need not lie between the two curves, and it will not be shown in other figures.
The two solid curves will be reused in subsequent figures with $\alpha' = \alpha$, in which case they form the standard symmetric confidence belt for the confidence level $\beta = 1 - 2\alpha$.
Smaller $\beta$ means a more narrow belt.

The following relation holds:

$$P\left(L_\alpha(\theta) < \hat{\theta} < U_{\alpha'}(\theta)\right) = 1 - \alpha - \alpha' \equiv \beta \tag{3}$$

Then an equivalent expression

$$P\left(u_{\alpha'}(\hat{\theta}) < \theta < l_\alpha(\hat{\theta})\right) = \beta \tag{4}$$

says that the probability for the random interval $[u_{\alpha'}(\hat{\theta}), l_\alpha(\hat{\theta})]$ to cover the unknown $\theta$ is $\beta$ (the confidence level; e.g. $\beta = 95\%$ etc.).

Choosing $\alpha = \alpha' = (1-\beta)/2$ results in a standard symmetric confidence belt.

The construction allows further freedom. Fix the confidence level $\beta$ (e.g. $\beta = 95\%$; $\beta$ is assumed to be fixed in what follows). Then choose a pair of functions $L$, $U$ to satisfy

$$P\left(L(\theta) < \hat{\theta} < U(\theta)\right) = \beta \tag{5}$$

If they are also monotonic then there exist inversions $u = U^{-1}$, $l = L^{-1}$, and an equivalent expression

$$P\left(u(\hat{\theta}) < \theta < l(\hat{\theta})\right) = \beta \tag{6}$$

says that the random interval $[u(\hat{\theta}), l(\hat{\theta})]$ covers the unknown $\theta$ with probability $\beta$.

Note that the curve $\theta = u(\hat{\theta})$ cannot exceed $\theta = u_{1-\beta}(\hat{\theta})$, i.e. $u(\hat{\theta}) \leq u_{1-\beta}(\hat{\theta})$. The curve $\theta = l(\hat{\theta})$ is similarly bounded from below.

Any such pair of curves forms what we will call *allowed confidence belt* for the confidence level $\beta$.

Fig. 2 introduces, in addition to the symmetric confidence belt for the confidence level $\beta$, another, narrower symmetric confidence belt for a lower confidence level $\tilde{\beta}=1-2(1-\beta)=1-4\alpha<\beta$ (dashed sloping lines). The various intersection points and horizontal lines are labeled for ease of reference: similarly labeled points in subsequent figures are the same as in this one.

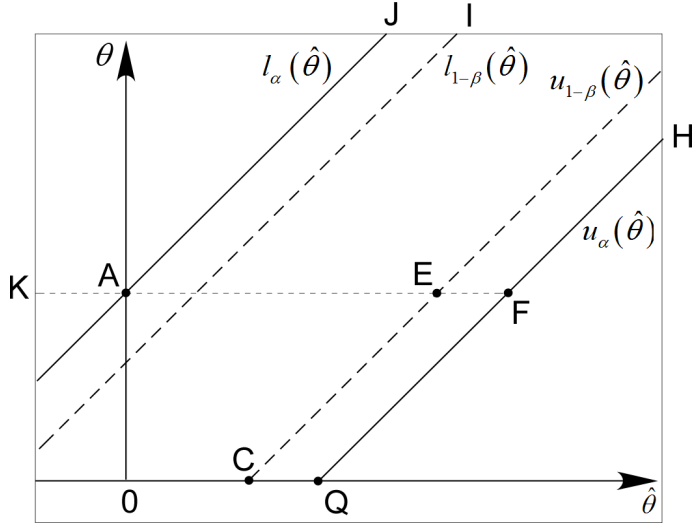

**Fig. 2.** The pairs of solid and dashed sloping lines delimit symmetric confidence belts for the confidence limits $\beta = 1-2\alpha$ and $\tilde{\beta}=1-2(1-\beta)=1-4\alpha$; cf. Fig. 1. The functions that correspond to the lines are shown in the figure. A is the intersection with the vertical axis of the line $\theta = l_\alpha(\hat\theta)$. Point A determines the horizontal line KF along with the further intersection points. C and Q are the intersections of the lines $\theta = u_{1-\beta}(\hat\theta)$ and $\theta = u_\alpha(\hat\theta)$ with the horizontal axis.

The vertical position of the intersection point A (and of the line KF) is denoted as $\theta_A$:

$$\theta_A = l_\alpha(0) \tag{7}$$

The numbers $\theta_C < \theta_E < \theta_F$ are the horizontal positions of the points C, E, F:

$$\theta_C = U_{1-\beta}(0), \quad \theta_E = U_{1-\beta}(\theta_A), \quad \theta_F = U_\alpha(\theta_A) \tag{8}$$

## 2. Horizontal deformations

The trick of horizontal deformations is based on the following properties of allowed confidence belts for a fixed $\beta$. If, for a fixed $\theta$, $U(\theta)$ in the definition (5) is pushed down to its lower limit $U_{1-\beta}(\theta)$ then the corresponding $L(\theta) \to -\infty$ ($L$ can be similarly pushed to its upper limit.) If $L$ and $U$ can thus be deformed while always preserving continuity and monotonicity, then there will be correctly defined inversions $u = U^{-1}, l = L^{-1}$ at every step of the deformation — i.e. an allowed confidence belt for the level $\beta$.

Given the way we draw plots in this letter, such deformations occur in horizontal directions. Fig. 3 illustrates this: the fat curves must lie outside the internal belt and may only approach its boundaries (in the horizontal directions shown by the arrows) on one side at the cost of running away from it to infinity on the other side.

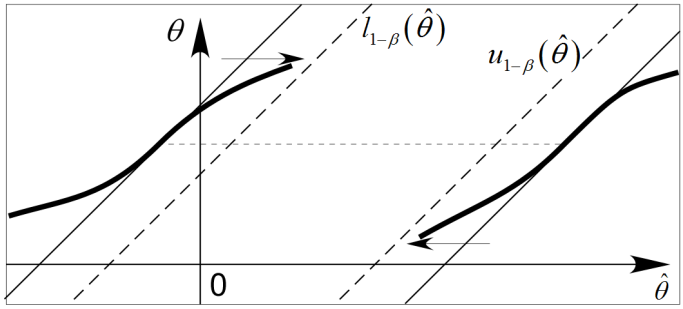

**Fig. 3.** The pairs of solid and dashed sloping curves delimit symmetric confidence belts for the confidence limits $\beta = 1 - 2\alpha$ and $\tilde\beta = 1 - 2(1-\beta) = 1 - 4\alpha$ (cf. Fig. 1). The two fat curves show a possible choice of $l$, $u$ for the confidence level $\beta = 1 - 2\alpha$. The arrows show horizontal deformations discussed in the main text.

Whenever one of the fat curves crosses a boundary of the symmetric confidence belt (the solid sloping lines) then the other fat curve crosses the other boundary, as shown with the horizontal dashed line.

The described freedom was exploited in ref. [2] where the pair $L$, $U$ was chosen based on a learned belief in the magic of likelihood. The choice of [2] is illustrated in Fig. 4.

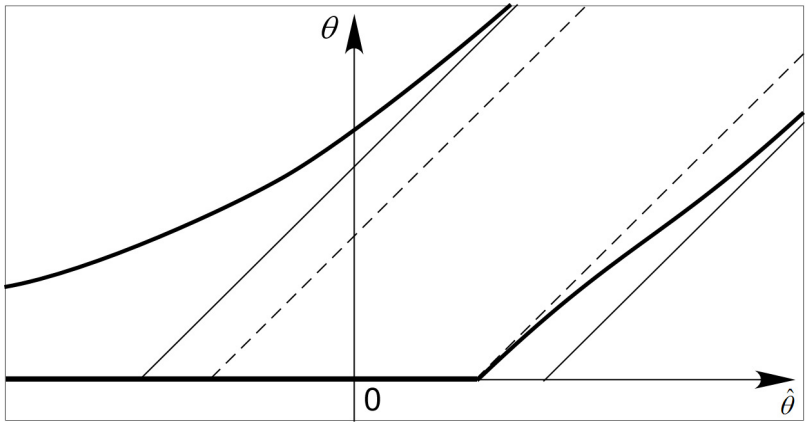

**Fig. 4.** The confidence belt (delimited by fat lines) as defined in ref. [2] (cf. their Fig. 10). The other lines are the same as in Fig. 3. The right fat curve approaching the dashed boundary corresponds to the left fat curve running away to infinity while approaching the horizontal axis, cf. Fig. 3.

Lastly, one can take a limit deforming $L$ so that its part adheres to a part of $L_{1-\beta}(\theta)$ (see Fig. 5 below). This may cause $L$ (and $U$) to loose continuity at the boundary of such part. However, if the corresponding inversions $u = U^{-1}$, $l = L^{-1}$ continuously approach, in the limit, well-defined continuous monotonic (non-decreasing) functions of $\hat\theta$, then the system of confidence intervals (6) will continuously approach a well-defined result, and the limiting confidence belt will be as good as any allowed confidence belt for the purposes of parameter estimation.

We are ready to go.

## 3. Taking into account the inequality $\theta \geq 0$

The defining element of the construction of confidence intervals is the estimator. What then should the estimator be instead of $\hat\theta = \hat\theta(X)$ if one knows beforehand that $\theta \geq 0$? The purpose of any estimator is to provide a value as close as possible to the unknown $\theta$. So, define a new estimator:

$$\tilde\theta = \max(\hat\theta, 0) \qquad (9)$$

Evidently, $\tilde\theta$ yields estimates that are guaranteed to be closer to the unknown value of $\theta$ than $\hat\theta$, and it incorporates both the statistical information contained in the unmodified estimator $\hat\theta$ as well as the a priori knowledge that $\theta \geq 0$. It then remains to construct confidence intervals for the new estimator $\tilde\theta$.

One may wish to ponder the definition (9) prior to reading on.

The probability distribution for $\tilde{\theta}$ has the form

$$\tilde{d}_\theta(\tilde{\theta}) = H(\tilde{\theta}) d_\theta(\tilde{\theta}) + c_\theta \delta(\tilde{\theta}) \tag{10}$$

where $H(t)$ is the standard Heaviside step function, $\delta(t)$ is the usual Dirac $\delta$-function and

$$c_\theta = \int_{-\infty}^{0} d\hat{\theta}\, d_\theta(\hat{\theta}) \tag{11}$$

So, one has to deal with the aggravation of a singular contribution. This can be done in a regular way, or via a trick. The trick is shorter and we will resume its description after first saying a few words about the regular way. Pragmatic readers may wish to skip the indented fragment.

> The regular method to deal with singular (generalized) functions that is rooted in their mathematical definition is to first approximate them by conventional ones with the approximation controlled by a small parameter, say $\varepsilon$, and to take the limit $\varepsilon \to 0$ in the end (this idea was first formulated by Fock [5]; the trick is sometimes called regularization). In our specific case, one has to choose a regularization for $\tilde{d}_\theta(\tilde{\theta})$ and construct the confidence intervals according to Neyman's prescription for each $\varepsilon$ in such a way that the confidence intervals obtained in the limit $\varepsilon \to 0$ not depend on the choice of the regularization. It is sufficient to define
>
> $$\tilde{\theta}_\varepsilon = \text{IF } \hat{\theta} \geq \varepsilon \text{ THEN } \hat{\theta} \text{ ELSE } \frac{\varepsilon}{2-\hat{\theta}/\varepsilon} \text{ END} \tag{12}$$
>
> This is a one-to-one mapping $(-\infty, +\varepsilon) \to (0, +\varepsilon)$ instead of $(-\infty, 0) \to 0$. Heuristically speaking, this means that negative instances of $\hat{\theta}$ are moved to the segment $(0, +\varepsilon)$. With the estimator thus redefined, one can, for each $\varepsilon$, construct confidence belts in a conventional manner. Then one takes the limit $\varepsilon \to 0$. While doing so, one must ensure that the result is independent of the particular form of the mapping (12).

Now, the trick. The key observation is as follows. The definition (9) means that the negative values of the unmodified $\hat{\theta}$ are carried over to the zero point and piled up there. This means that all such values will be indistinguishable: they will all yield the zero value for $\tilde{\theta}$ — and the same confidence interval. This implies that all non-positive values of $\hat{\theta}$ are to eventually yield one and the same confidence interval [0, const], where the constant is independent of $\hat{\theta}$.

Once this is understood, the construction of confidence belts can be completed entirely in terms of $\hat{\theta}$, with the only trace of $\tilde{\theta}$ being an **additional condition**: the resulting confidence belts must be such that all values of $\hat{\theta}$ below the a priori bound must yield the same confidence interval.

The condition has a clear experimental meaning: it can be rephrased as a requirement of robustness of confidence intervals with respect to unknown experimental (ef|de)fects. This gives the entire construction an additional physical weight. However, it is worth keeping in mind that the argument that led to it starting from the beginning of this section is transparent and specific, and does per se not need any metaphysical support: the construction in terms of the unmodified estimator $\hat{\theta}$ with the above additional condition is equivalent to a straightforward construction of confidence belts for the modified estimator (9) that incorporates the a priori information in a most straightforward and transparent fashion.

Our construction will employ the device of horizontal deformations described in Sec. 2.

In the notations of Fig. 2 and with a fixed confidence level $\beta$, consider a confidence belt corresponding to two functions $l$, $u$ chosen as shown in Fig. 5 (cf. also Figs. 3 and 4).

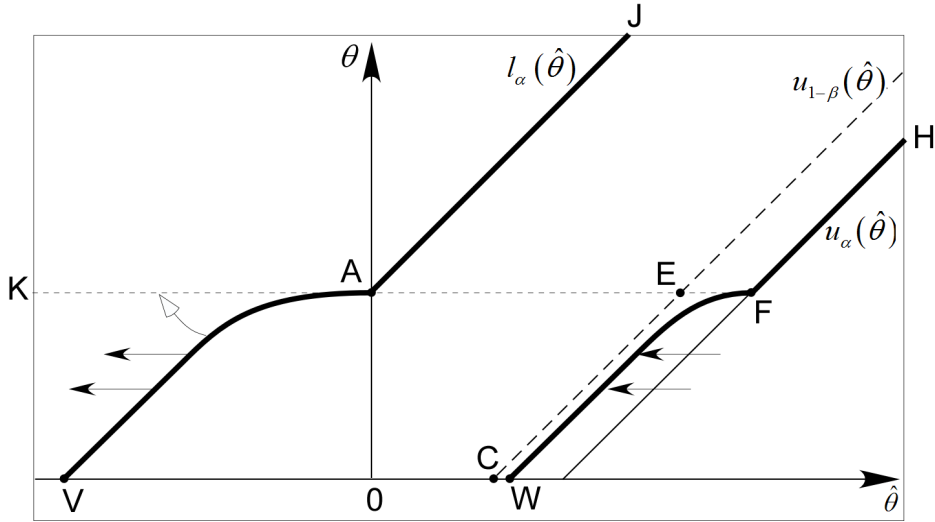

**Fig. 5.** Fat lines delimit an allowed confidence belt for $\hat\theta$ for the confidence level $\beta$. The other lines and points are as in Fig. 2. The fat lines are described by two functions $l$ and $u$. Black arrows indicate the horizontal deformation used to obtain the confidence belt that satisfies the additional condition.

As was discussed in Sec. 2, one is allowed to perform the horizontal deformation indicated by the black arrows in Fig. 5 until the curved segment WF adheres to the broken line CEF. The curved segment AV will in the end effectively be straightened out into AK (the white arrow). The limiting confidence belt is well-defined and is shown in Fig. 6.

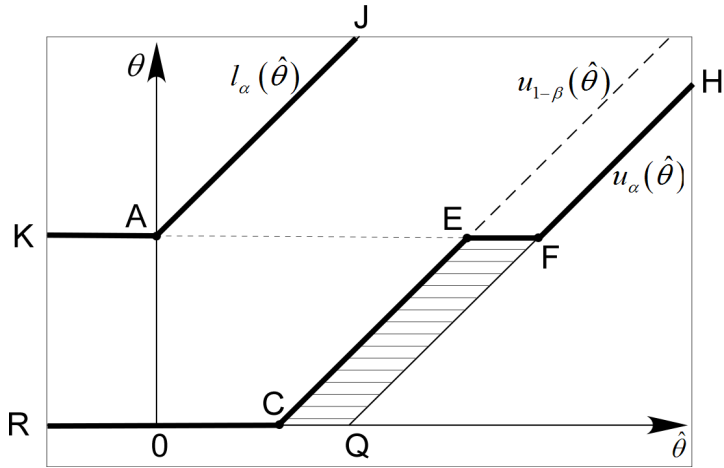

**Fig. 6.** The two fat broken lines KAJ and RCEFH delimit the resulting level-$\beta$ confidence belt for $\hat\theta$ that satisfies the additional condition and therefore correctly maps to a confidence belt for $\tilde\theta$ defined by eq. (9).
The region CEFQ cut out from the unmodified confidence belt is a pure gain obtained from the a priori knowledge.

Recall that the construction involves two symmetric confidence belts for the unmodified $\hat\theta$:

1) the symmetric belt for the level $\beta = 1 - 2\alpha$, expressed in our notation as $\left[u_\alpha(\hat\theta), l_\alpha(\hat\theta)\right]$,

2) the symmetric belt for the level $\tilde\beta = 1 - 2(1-\beta) = 1 - 4\alpha$, expressed as $\left[u_{1-\beta}(\hat\theta), l_{1-\beta}(\hat\theta)\right]$.

The resulting confidence belt for the confidence level $\beta$ is analytically described as follows (the various $\theta_X$ are defined in eqs. (7) and (8)):

— For $\hat{\theta} \geq \theta_F$, the confidence interval is the standard symmetric interval for $\hat{\theta}$ for the confidence level $\beta = 1 - 2\alpha$, as if the parameter were not bounded from below, i.e. $[u_\alpha(\hat{\theta}), l_\alpha(\hat{\theta})]$.

— For $\theta_E \leq \hat{\theta} \leq \theta_F$, the confidence interval is $[\theta_A, l_\alpha(\hat{\theta})]$, i.e. the upper bound as before, and the lower bound $\theta_A$.

— For $\theta_C \leq \hat{\theta} \leq \theta_E$, the confidence interval is $[u_{1-\beta}(\hat{\theta}), l_\alpha(\hat{\theta})]$, i.e. the upper bound as before, and the lower bound the same as for the standard symmetric interval for the confidence level $\tilde{\beta} = 1 - 2(1-\beta) = 1 - 4\alpha$.

— For $0 \leq \hat{\theta} \leq \theta_C$, the confidence interval is $[0, l_\alpha(\hat{\theta})]$, i.e. the upper bound as before, and the lower bound 0.

— Lastly, for all $\hat{\theta} \leq 0$, the confidence interval is fixed as $[0, \theta_A]$.

The noteworthy properties of this confidence belt are as follows:

♥ The estimate is robust for the non-physical values of the estimator, i.e. for $\hat{\theta} < 0$.

♥ The interval's upper bound for the physical values of the estimator is the same as in the unmodified case, and is always lower than that of ref. [2].

♥ The interval's lower bound breaks off zero at the earliest point possible for the confidence level $\beta$ (at $\theta_C$), and the lower bound is maximal possible for any allowed confidence intervals at this confidence level, in the interval $\theta_C \leq \hat{\theta} \leq \theta_E$.

♥ Neither complicated algorithms nor tables are required on top of the standard routines to compute conventional confidence intervals for various confidence levels.

## 4. Conclusions

The key point of our construction is the definition (9). One can consider other modifications of the estimator, e.g. $\breve{\theta} = |\hat{\theta}|$. However, such alternative modifications lead to suboptimal resolution of non-zero values of the parameter (somewhat similar to the construction of [2], cf. Fig. 4). One also has to remember that the difficulty addressed by this analysis is encountered in measurements where the unknown $\theta$ is expected to lie close to the a priori bound (zero in our notations), so that in the absence of further specific a priori information about the unknown $\theta$, eq. (9) is the optimal choice.

To conclude, the prescription represented by Fig. 6 (the analytical description is given after the figures) solves the problem of incorporating a priori information in the form of a bound $\theta \geq 0$ into the construction of confidence belt in a simple, general, purely frequentistic, and fully motivated fashion.
The obtained prescription (unlike the one given in ref. [2]) is optimal in the sense that it yields the best and earliest resolution of non-zero values from zero while being robust in the unphysical region. It is also quite general and is constructed from pieces of two conventional confidence belts for confidence levels $\beta$ and $\tilde{\beta} = 1 - 2(1-\beta) < \beta$.

Notes added (2011-04-05).

The Mainz neutrino mass experiment [6] mentioned the so-called *sensitivity limit* as a way to present an upper bound for the case of a negative estimate for an a priori non-negative parameter (mass squared in their case). Such a sensitivity limit exactly corresponds to the KA part of our Fig. 6. However, in the final presentation of their measurement, they switched to the unified approach estimate of [2] as recommended by PDG at the time. The present work can be regarded as providing a systematic justification for the sensitivity limit upper bound for non-physical values of the estimator, while also providing the maximal tightening of the lower part of the confidence belt in the physical region (the CEF part in our Fig. 6).

Ref. [7] explored, for completeness's sake, a modification of the maximally asymmetric confidence intervals (bounds) in presence of a priori bound using the techniques of the present paper. However, for the purpose of reporting an upper bound, the method of the present paper seems preferable as it provides a proper connection with the symmetric confidence belt in the physical region.

*Acknowledgements.* A.S. Barabash (of the double beta-decay) impressed upon me the importance of the problem by relating the controversies around data processing in small-signal experiments. Yu.M. Andreev (of CMS) pointed out ref. [2] specifically but I failed to appreciate it until A.A. Nozik (of the Troitsk ν-mass) raised the issue in a specific context, brandishing a Bayesian argument. A.V. Lokhov, G.I. Rubtsov and V.I. Umatov lent ear to early versions of this account, thus helping to improve it. Finally, I thank the members of the Troitsk ν-mass experiment for creaing a stimulating context for this work.

## References

[1] Troitsk ν-mass: http://www.inr.ru/~trdat/
KATRIN: http://www-ik.fzk.de/tritium/motivation/index.html

[2] G.J. Feldman and R.D. Cousins. A unified approach to the classical statistical analysis of small signals. arXiv:physics/9711021.

[3] W.T. Eadie, D. Dryard, F.E. James, M. Roos and B. Sadoulet. Statistical methods in experimental physics. North-Holland, 1971.

[4] F.V. Tkachov. Transcending the least squares. arXiv:physics/0604127

[5] V.A. Fock. On "improper" functions in quantum mechanics. J. of Russian Physical and Chemical Society, 61, no.1 (1929) 1 (in Russian); http://www.inr.ac.ru/~ftkachov/misc/Fock1929/
L. Schwartz. Theorie des distributions. Vol. 1. Hermann, 1950.
I.M. Gelfand and G.M. Shilov. Generalized functions. Vol. 1. Moscow, 1959 (in Russian).
L. Schwartz. Methodes mathematiques pour les sciences physiques. Hermann, 1961.
M. Reed and B. Simon. Methods of modern mathematical physics. Vol. 1. Academic, 1972.
P. Antosik, J. Mikusinski and R. Sikorski. Theory of distributions. Elsevier, 1973.
R.D. Richtmyer. Principles of advanced mathematical physics. Vol. 1. Springer-Verlag, 1978.

[6] Ch. Kraus et al. Final results from phase II of the Mainz neutrino mass search in tritium $β$ decay. arXiv:hep-ex/0412056v2

[7] F.V. Tkachov. Optimal upper bounds for non-negative parameters. arXiv:0912.1555.